\documentclass{optica-article}

\journal{opticajournal} 

\articletype{Research Article}

\usepackage{siunitx}

\begin{document}

\title{Spatiotemporal wavevector dynamics of transverse mode-locked beams}
%

\author{Jan Wichmann\authormark{*}, Michael Zwilich, and Carsten Fallnich}

\address{University of Münster, Institute of Applied Physics, Correnstr. 2, 48149 Münster, Germany}

\email{\authormark{*}jan.wichmann@uni-muenster.de} 


\begin{abstract*} 
The phase-locking of transverse higher-order resonator modes results in a beam with rapidly oscillating spatial intensity profile. 
To complete the description of such transverse mode-locked (TML) beams, their spatiotemporal phase dynamics are explored in this work. The analysis of the phase terms of TML beams reveals oscillating wavevectors, which are experimentally verified by exploiting the mode-matching requirement of a single-mode fiber. The wavevector oscillation is further quantified by exploring its dependence on beam size. The findings have to be considered in potential future applications of TML beams in phase-sensitive processes.
\end{abstract*}


\section{Introduction}
Mode-locking is the fixed-phase superposition of resonator modes, where longitudinal mode-locking (LML) of  resonator modes can result in ultrashort laser pulses. 
In contrast, a transverse mode-locked (TML) beam resulting from a fixed-phase superposition of higher-order transverse laser modes was shown to exhibit spatiotemporal dynamics of the intensity distribution, i.e., a laser beam that periodically scans along the transverse axis \cite{Auston1968,Auston1968a}.
TML light fields do expand the group of spatiotemporally coupled light fields \cite{Wright2017,Shen2023}, with which spatial features cannot be described separately from their time evolution \cite{Pariente2015}.
In other words, for spatiotemporally coupled light fields, the complex spatial field distribution changes with time, but not only by a time-dependent factor that affects the entire spatial distribution identically. Instead each point in its spatial distribution undergoes its own time evolution, meaning the field changes its shape with time. 
This correspondence between a spatial feature and a point in time in principle allows for spatial information to be drawn from a time-resolved measurement and vice versa.

Scanning laser beams are used in a range of imaging \cite{Ra2007,Ward2017}, materials processing \cite{Kraetzsch2011,Prieto2020}, and medical applications \cite{Huang2020,ChanghoChong2006}, and different scanning rates ranging from a few $\mathrm{Hz}$ to a few $\mathrm{MHz}$ are used depending on the application. Many applications of scanning laser beams would benefit from higher scanning rates, since they could allow for shorter processing times.
For TML beams, rapid scanning rates are feasible because their repetition rate is given by the resonator geometry instead of moving optical components, like galvanometric mirrors, which would limit scanning speeds due to inertia. Previous works on TML beams have shown rapid scanning rates of up to $\qty{90}{MHz}$ \cite{Auston1968,Schepers2020,Zwilich2024}. With as of yet unrealized passive TML, even faster scanning rates beyond the $\mathrm{GHz}$-regime are theoretically feasible, because TML scanning rates are ultimately determined by the resonator geometry only.

The spatiotemporal intensity dynamics of TML beams have been successfully observed in a number of systems including free-space laser resonators \cite{Auston1968,Watanabe1968,Kohiyama1968,Ito1969,Haug1974,Schepers2020}, fiber lasers \cite{Wright2017}, laser diodes \cite{Logginov2004}, and by conversion of LML beams within empty resonators \cite{Zwilich2023,Zwilich2024}.
While many applications involving scanning laser beams only take advantage of the intensity dynamics of these beams, phase-sensitive processes like optical fiber coupling or nonlinear frequency conversion require an understanding of the phase dynamics, which have yet to be investigated for TML beams.
In the following, we explore the spatiotemporal phase dynamics of TML beams, where the analysis of the TML beam's spatial phase reveals a spatiotemporal wavevector oscillation. 
We experimentally verify this finding by exploiting the coupling condition of a single-mode optical fiber (SMF) \cite{Wallner2002}, where the input field and SMF eigenmode fields have to be maximally similar in their complex spatial field distributions to maximize transmission. Mode-matching thus includes also matching of the phase distributions of input and eigenmode field, making the SMF a viable phase-sensitive measurement device for the characterization of a TML beam. 
This investigation into the spatiotemporal phase dynamics of TML beams completes the description of their spatiotemporal evolution and has to be accounted for regarding potential applications.
\section{Wavevectors of transverse mode-locked light fields}
The TML beam under investigation here is constituted by a number of Hermite-Gaussian (HG) modes, which are usually identified by two mode indices $(m,n)$, each for one transverse axis. In the following, the mode index along the $x$-axis is set to correspond to the fundamental mode ($m=0$), to constrain the higher-order modes and therefore the resulting beam scanning to the $y$-axis. Furthermore, the beams are investigated in the beam waist resulting in a plane wave phase factor $\exp{\left[i(k_{z,n}z-\omega_{n}t)\right]}$.
The HG resonator modes are then given by
\begin{equation}
    \mathrm{HG}_{n}\equiv\mathrm{HG}_{0,n}(\xi,z,t)=\frac{1}{N_{n}}H_{n}(\xi)\exp{\left[-\xi^{2}/2\right]}\exp{\left[i(k_{z,n}z-\omega_{n}t)\right]},
    \label{eq:modes}
\end{equation}
where $\xi=\sqrt{2}y/w_0$ is the normalized transverse coordinate, with the $1/e$-beam radius of the fundamental mode $w_0$, the resonance frequency $\omega_{n}$, its corresponding longitudinal wavevector $k_{zn}=\omega_{n}/c$, $c$ the vacuum velocity of light, normalizing factor $N_{n}=\sqrt{2^n n! \sqrt{\pi}}$, and $H_n(\xi)$ being the $n$-th order Hermite polynomial. The resonance frequencies of the higher-order transverse modes $\omega_{n}=\omega_{0}+n\Omega.$ are shifted from the fundamental mode frequency $\omega_0$ by the transverse mode spacing $\Omega$, which is due to the Gouy phase shift.

\begin{figure}[h]
\centering
\includegraphics[width=1\linewidth]{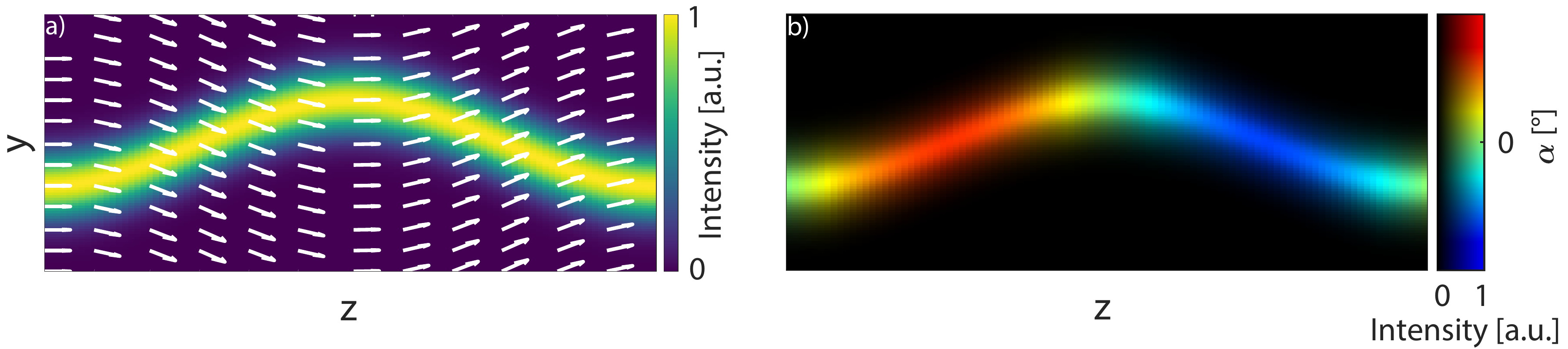}
\caption{\label{fig:k_vector}a) TML beam intensity within one oscillation period along $z$ with wavevectors according to Eq. (\ref{eq:wavevector}) at different points in space overlayed as arrows. b) For ease of readability a colorcoded representation of the wavevector angles is introduced. The angles corresponding to the arrows shown in a) are encoded as different colors, whereas intensity is encoded as brightness.}
\end{figure}

As the resonator modes $\mathrm{HG}_{n}(\xi,z,t)$ exhibit different resonance frequencies $\omega_{n}$, in superposition they display a spatiotemporal beating, such that both spatial intensity distribution as well as phase distribution vary with time. A superposition of the resonator modes (Eq. (\ref{eq:modes})) yields an electric field distribution
\begin{equation}
    E(\xi,z,t)=\sum_{n}^{\infty}A_{n} \exp{\left[i(k_{z,n}z-\omega_{n}t)\right]}\frac{1}{N_{n}}H_{n}(\xi)\exp{\left[-\xi^{2}/2\right]},
    \label{modes_superpos}
\end{equation}
where $A_{n}=\left|A_{n}\right|\exp{(i\phi_{n})}$ are the complex modal coefficients that encode the modal power share $\left|A_{n}\right|^2$ and the phase $\phi_{n}$ of each mode involved in the superposition.
When superimposing a set of these modes with Poisson-distributed modal powers and a linear modal phase relation, a Gaussian transverse spatial intensity distribution results, oscillating along the transverse $y$-axis with the frequency $\Omega$\cite{Auston1968} as shown in Fig. \ref{fig:k_vector} a). Such a Poissonian superposition of transverse modes with a linear modal phase will here be referred to as a $\mathrm{TML}^{(\mathrm{p})}$ beam, where the index $(\mathrm{p})$ stands for Poissonian. The $\mathrm{TML}^{(\mathrm{p})}$ beam's electric field superposition of the individual modes can be simplified to
\begin{align}
    E(\xi,z,t)=\pi^{-\frac{1}{4}}\exp[-1/2(\xi-\xi_{0}\cos{(Kz-\Omega t)})^{2}]\label{eq:TML}\\\nonumber\cdot\exp[i(k_{0}z-\omega_{0}t)]\exp[i\xi_{0}\sin{(Kz-\Omega t)}(\xi-\xi_{0}/2\cos{(Kz-\Omega t)})],
\end{align}
where $K=\Omega/c$ and the normalized spatial oscillation amplitude $\xi_{0}=\sqrt{2\Bar{n}}$ were introduced, with $\Bar{n}=\sum_{n} n|A_{n}|^2/\sum_{n} |A_{n}|^2$ being the average mode order.
The transverse spatial oscillation of the intensity distribution is described by the first exponential, which has been the primary focus in previous research regarding TML \cite{Auston1968, Schepers2020}. In this work instead, the spatiotemporal phase evolution, i.e., the following exponentials in the second line of Eq. (\ref{eq:TML}) are investigated:
\begin{equation}
    \theta_{\mathrm{total}}(\xi,z,t)=k_{0}z-\omega_{0}t+\xi_{0}\sin{(Kz-\Omega t)}(\xi-\xi_{0}/2\cos{(Kz-\Omega t)}).
\end{equation}
Using $\xi=\sqrt{2}y/w_0$ and the trigonometric identity $\sin{(\theta)}\cos{(\theta)}=\sin{(2\theta)}/2$, the wavevector can be calculated as the rate of phase change with respect to the involved spatial dimensions
\begin{equation}
\mathbf{k}=
    \begin{pmatrix} k_{y}  \\ k_{z} \end{pmatrix} = \begin{pmatrix} \partial \theta_{\mathrm{total}}/\partial y  \\ \partial \theta_{\mathrm{total}}/\partial z \end{pmatrix}=\begin{pmatrix} \xi_{0}\sqrt{2}\sin{(Kz-\Omega t)}/w_0 \\ k_{0}+K\xi_{0}\sqrt{2}y\cos{(Kz-\Omega t)}/w_0-K\xi_{0}^{2}\cos{(2(Kz-\Omega t))}\end{pmatrix}.
    \label{eq:wavevector}
\end{equation}
Note, that here $\partial w_0 /\partial z=0$, because the field is investigated in the beam waist.
The explicit dependence on time as well as space in Eq. (\ref{eq:wavevector}) highlights the tilting of the wavevector during the oscillation by an angle $\alpha=\arctan{\left({k_{y}}/{k_{z}}\right)}$.

As shown by horizontal arrows in Fig. \ref{fig:k_vector} a), the wavevector is collinear with the optical axis $z$ only at the return points of the beam, where the beam is maximally displaced from its center. In contrast, within the central sections, where the beam moves between the return points, the wavevector exhibits nonzero angles to the optical axis. The direction of the wavevector appears highly unintuitive, as it does not seem to point along the propagation direction of the beam except at the outer return points. A more intuitive understanding can be gained from the time- and space-dependent Visualization 1 as provided in the supplementary material.\\
From Eq. (\ref{eq:wavevector}) follows that the maximum wavevector tilt occurs at $Kz-\Omega t= \pi/2$ with a magnitude of
\begin{equation}
\alpha_{\mathrm{max}}
=\arctan{\left(\frac{\xi_{0}\sqrt{2}/w_0}{k_{0}+K\xi_{0}^{2}}\right)}.
    \label{eq:max_alpha}
\end{equation}
Thus, $\alpha_{\mathrm{max}}$ increases with decreasing fundamental mode radius $w_0$ or increasing normalized spatial oscillation amplitude $\xi_0=\sqrt{2\Bar{n}}$, approaching $\alpha_{\mathrm{max}}=\pi/2$ as its theoretical limit for $w_0\to0$.\\
The predicted wavevector tilt has significant implications for many processes that require phase-matching, such as nonlinear frequency conversion or optical fiber coupling.
The coupling efficiency of an input field to an SMF eigenmode is determined by the overlap integral of their complex field distributions:
\begin{equation}
    \eta=\frac{\left|\int \left(E_{\mathrm{in}}(y)e^{i\psi(y)}\right)^{*}E_{\mathrm{mode}}(y)dy\right|^{2}}{\int\left|E_{\mathrm{in}}(y)\right|^2dy \int\left|E_{\mathrm{mode}}(y)\right|^2dy}.
    \label{eq:coupling}
\end{equation}
According to Eq. \ref{eq:coupling}, the input light field $E_{\mathrm{in}}$ has to be matched to the SMF eigenmode $E_{\mathrm{mode}}$, which is here assumed to be a Gaussian fundamental mode, to achieve maximum coupling efficiency. A wavevector angle $\alpha$ of the input field relative to the axis of the SMF introduces a phase factor $e^{i\psi(y)}$, which reduces the overlap of input and eigenmode fields and thus the coupling efficiency. The additional phase introduced through the mismatch of wavevector angle and fiber angle is given by $\psi(y)=k_{y}y=|\mathbf{k}|\sin{(\alpha)}y$.
The overlap of the input field $E_{\mathrm{in}}$ and the SMF eigenmode $E_{\mathrm{mode}}$ is maximized when $E_{\mathrm{in}}=E_{\mathrm{mode}}$ and $\alpha=0$. The angular mismatch between fiber angle and wavevector angle can be compensated by orienting the fiber at an angle identical to the wavevector angle of the incoming beam, thus restoring maximum coupling efficiency.\\
When applying this general model of wavevector-angle dependent optical fiber coupling to a TML beam, it quickly becomes obvious how the coupling efficiency of different sections of the beam would be affected.
The tilted wavevectors at the central sections of the TML beam would diminish the coupling efficiency to the eigenmode of an SMF that is oriented parallel to the $z$-axis, while only the beam sections near the return points of the beam would efficiently couple to the SMF when it is aligned parallel to the $z$-axis.
In contrast, when orienting the fiber axis according to the wavevector angle at the central sections of the TML beam, the wavevector mismatch would decrease at those sections, allowing for increased coupling efficiency, while the mismatch would increase for sections at the return points of the TML beam, that were previously well-aligned to the fiber axis.
This wavevector-selective coupling efficiency makes the SMF a viable phase-sensitive tool to probe the TML beams spatiotemporal wavevector dynamics, i.e., by measuring the transmitted time-dependent intensities at different fiber positions and angles.


\section{Experimental setup and results}
The TML laser used in our experiments was a plano-concave end-pumped $\mathrm{Nd\!:\!\!YVO_{4}}$ laser, where the transverse modes were actively locked by modulating the cavity loss with an acousto-optic modulator (AOM) with an effective frequency $\nu_{\mathrm{mod}}=\qty{82.2}{MHz}$ that matched the transverse mode frequency spacing $\Omega$\cite{Schepers2020}. Suppression of higher-order modes along the $x$-axis was ensured by introducing a slit into the cavity, allowing for generation of higher-order modes only along the $y$-axis.

\begin{figure}[h]
    \centering
    \includegraphics[width=1\linewidth]{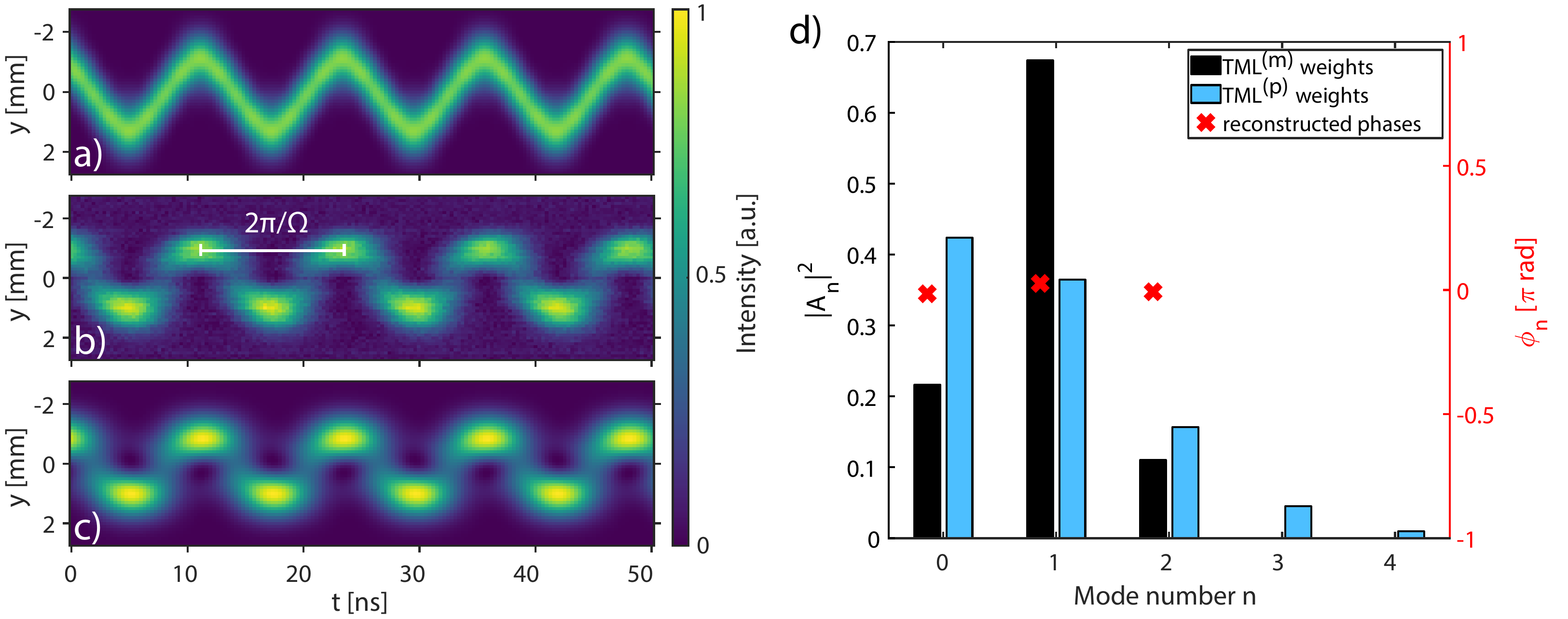}
    \caption{a) Spatiotemporal intensity distribution of a $\mathrm{TML}^{(\mathrm{p})}$ beam with $\Bar{n}=0.86$. b) Spatiotemporal intensity distribution recorded with a fast photodetector by scanning it across the generated TML beam along the $y$-axis. c) Reconstructed spatiotemporal intensity distribution based on the data of b). d) Reconstructed modal weights and phases for the measured $\mathrm{TML}^{(\mathrm{m})}$ beam and a $\mathrm{TML}^{(\mathrm{p})}$ beam.}
    \label{fig:beam_charac}
\end{figure}
\begin{figure}[h]
\centering
\includegraphics[width=0.75\linewidth]{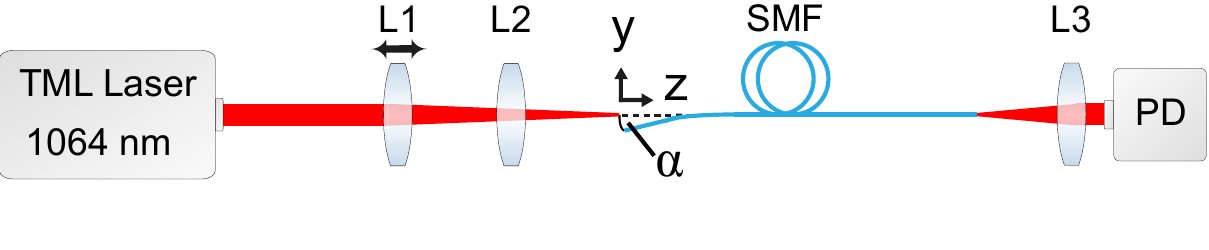}
\caption{\label{fig:setup}Schematic of the experimental setup. The beam from the TML laser is focussed by a combination of lenses (L1, L2), with the first one being placed on a linear shift to adjust the beam size at the SMF facet. The SMF can be translated along the oscillation ($y$-) axis of the beam and rotated in the $yz$-plane, resulting in fiber angles of $\alpha$ to the optical ($z$-) axis. Additionally, the change in $z$-position of the focal plane due to the variable lens combination (L1, L2) can be compensated for by $z$-shifting the fiber facet.}
\end{figure}

The spatiotemporal intensity distribution of the generated TML beam was characterized by scanning a fast photodetector ({Femto HSA-X-S-1G4-SI}, $250\,\mathrm{ps}$ rise time) along the $y$-axis  and reconstructing the beams modal weights and phases with a stochastic parallel gradient descent (SPGD) algorithm as in reference \cite{Schepers2020}. The TML resonator was adjusted to produce a modal weight distribution as shown in Fig. \ref{fig:beam_charac} d), with an average mode order of $\Bar{n}=0.86$, since larger $\Bar{n}$ would lead to larger unnormalized spatial oscillation amplitudes $Y_0=\xi_0 w_0/\sqrt{2}=\sqrt{\Bar{n}}w_0$ and according beam clipping at the lens apertures.
The Pearson correlation coefficient \cite{Schepers2020,Zwilich2023} between measurement and reconstruction was $0.98$, demonstrating good agreement, as can also be seen by comparing Fig. \ref{fig:beam_charac} b) and c).
The reconstructed beam's modal phase as deduced by the SPGD algorithm (Fig. \ref{fig:beam_charac} d)) was flat as for a $\mathrm{TML}^{(\mathrm{p})}$ beam (not shown), but the modal weights deviated by an average of $12 \%$ of total power from those of the $\mathrm{TML}^{(\mathrm{p})}$ beam, leading to a deviation in the spatiotemporal intensity distribution (cf. Fig. \ref{fig:beam_charac} a) and c)). Despite this, the spatial intensity scanning dynamics were still qualitatively present in the measured TML beam, such that the predicted wavevector dynamics should be contained as well. The measured TML beam with modal weights and phases as shown in Fig. \ref{fig:beam_charac} d) will be referred to as a $\mathrm{TML}^{(\mathrm{m})}$ beam. The oscillation frequency of the TML beam was identical to the AOM modulation frequency $\Omega/(2\pi)=\nu_{\mathrm{mod}}=\qty{82.2}{MHz}$.
After aligning the resonator to produce the desired TML beam, this beam exited the laser and passed a combination of lenses (L1: $f=\qty{100}{mm}$, L2: $f=\qty{3.1}{mm}$, aspheric, $\mathrm{NA}=0.68$) with variable separation (Fig. \ref{fig:setup}), allowing for an adjustable beam width at the effective focal plane, where it encountered the fiber facet of a {Corning Hi1060 Flex} SMF (mode field radius $w_{\mathrm{MFR}}=\qty{2}{\um}$ at $\lambda=\qty{980}{nm}$ wavelength). A $z$-shift in focal plane position introduced by the variable lens combination (L1, L2) was compensated by translation of the fiber facet position along the $z$-axis. The fiber facet could also be translated along the $y$-axis to fiber facet positions $\Tilde{y}$ and oriented at angles $\alpha$ to the propagation axis of the beam using computer-controlled translation and rotation mounts, respectively.
Upon coupling to the fiber, the coupled light field was guided in the fiber for a length of $\qty{40}{cm}$ and was then coupled out using an aspheric lens (L3: $f=\qty{6}{mm}$, $\mathrm{NA}=0.4$). The fiber length was sufficient for damping of cladding modes \cite{Wallner2002}, that would otherwise compromise the spatial and wavevector filter action of the fiber, while still minimizing loss for an improved signal-to-noise ratio at the detection unit. The transmitted light field was detected with a fast photodetector (PD, {Femto HSA-X-S-1G4-SI}, $\qty{250}{ps}$ rise time) and the resulting time signal was recorded with an oscilloscope (AGILENT Infiniium DSO80204B, $\qty{2}{GHz}$ bandwidth).
The focal length of the lens combination was chosen such that the fundamental mode radius $w_0$ was larger, but close to the SMF mode field radius $w_{\mathrm{MFR}}$ for all measurements in order to achieve sufficient transmission and because the non-normalized spatial oscillation amplitude $Y_0=\sqrt{\Bar{n}}w_0\approx\qty{5}{\um}$ was then also larger than $w_{\mathrm{MFR}}$, allowing for selective probing of the TML beam sections. This means, for a given fiber facet position only a section of the beam's spatial distribution could be coupled into the fiber. By moving the fiber along the oscillation axis of the beam, the TML beam's coupling to the SMF eigenmode could be probed for different positions $\Tilde{y}$ and angles $\alpha$.

For each angle of the fiber, the fiber facet was scanned completely across the TML beam along its oscillation axis ($y$-axis) to map the transmission properties for each angle to the different fiber positions. This way, parts of the beam could be selectively coupled to the fiber, dependent on their position and wavevector direction. Optimal coupling occured only at a given time if both beam position and wavevector were  matched by fiber facet position and fiber rotation, respectively. Due to the spatial scanning of the beam and the spatial filtering action of the fiber the beam sections were measured as pulse trains at the photodiode, with $1/e$-pulse durations around $\tau=\qty{2.3}{ns}$.\\
\begin{figure}[h]
\centering
\includegraphics[width=0.9\linewidth]{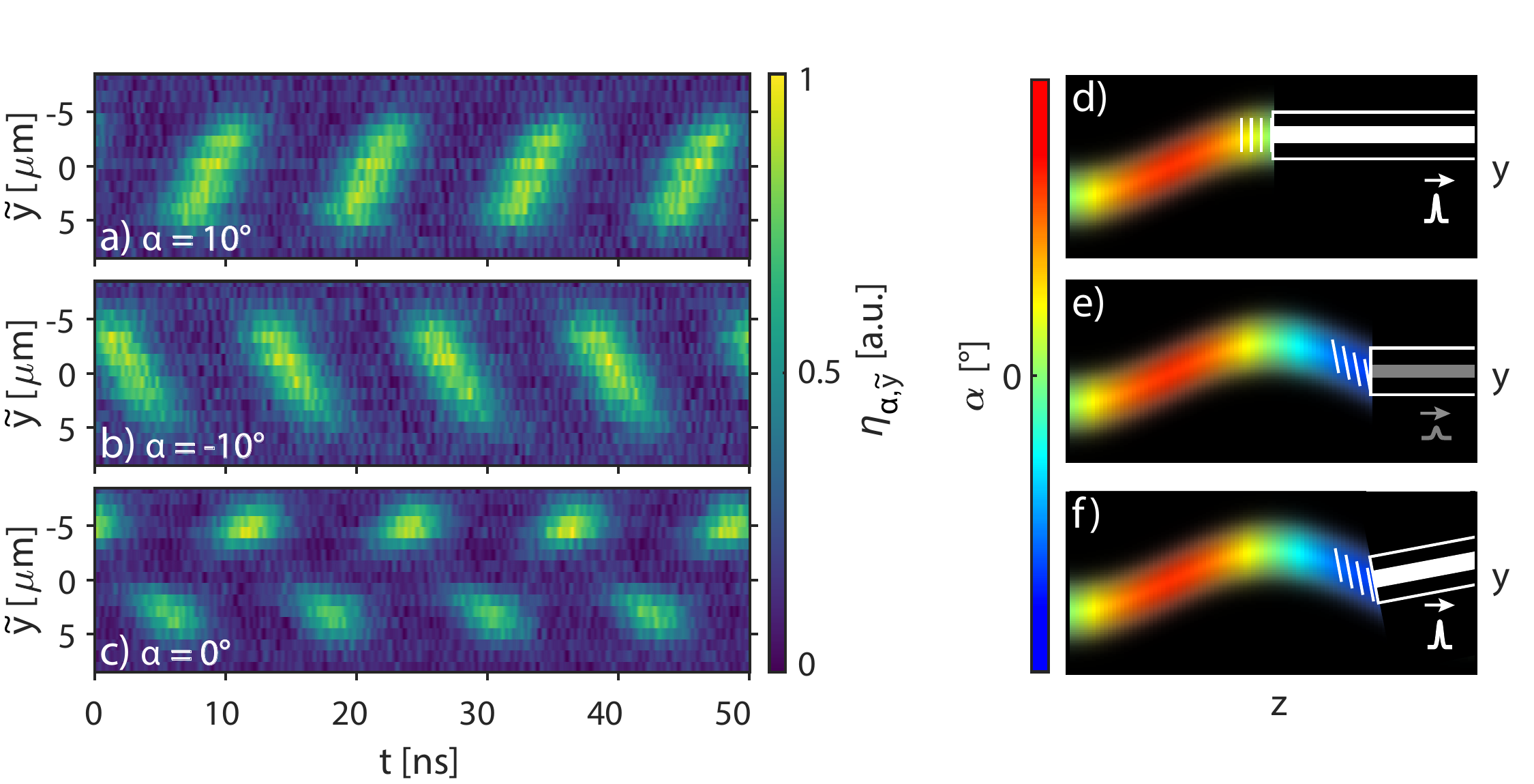}
\caption{\label{fig:yscans} a)-c) Exemplary photodiode time-traces for different positions $\Tilde{y}$ of the fiber facet and different fiber angles with a) $\alpha=10 ^{\circ}$, b) $\alpha=-10 ^{\circ}$ and c) $\alpha=0 ^{\circ}$. In d)-f) different cases for fiber coupling are illustrated. The spatially oscillating TML beam is shown with its intensity coded as brightness and its wavevector angles colorcoded.  Spatiotemporal beam sections that couple to the fiber are shown as pulses that are transmitted through the SMF, with more intense pulses for well-matched wavevectors and otherwise less intense pulses.}
\end{figure}
With the computer-controlled rotational mount the fiber was oriented at angles $-15^{\circ}<\alpha<+15^{\circ}$, because no signal was detectable for angles beyond that range. The angular range was scanned in integer degree steps to maintain accuracy within a reasonable measurement duration with an angular positioning error of $\pm 0.1^{\circ}$.
For each fiber angle, the fiber facet was scanned across the y-axis while recording a photodiode trace for each $\Tilde{y}$-position of the fiber facet, resulting in measurements like the ones shown exemplarily in Fig. \ref{fig:yscans} a)-c). When the fiber was oriented at an angle to the optical axis ($\alpha=\pm 10^{\circ}$, Fig. \ref{fig:yscans} a)-b)), the center of the beam ($\Tilde{y}=\qty{0}{\mu m}$) could be transmitted strongly, since the fiber was aligned with the wavevector of the TML beam (Fig. \ref{fig:yscans} f)), while the return points of the beam coupled to the fiber only weakly due to wave vector mismatch. When the fiber was oriented parallel to the optical axis ($\alpha=0^{\circ}$, Fig. \ref{fig:yscans} c)) instead, sections of the beam were transmitted only if the fiber facet was positioned at the return points of the TML beam at $\Tilde{y}\approx\pm \qty{5}{\um}$, as sketched in Fig. \ref{fig:yscans} d), while the beam center was not transmitted due to wave vector mismatch (Fig. \ref{fig:yscans} e)). In accordance to the time-periodicity of the TML beam, the detected transmission signal was periodic with the frequency $\Omega/(2\pi)=\qty{82.2}{MHz}$ for both instances as well.
Fig. \ref{fig:yscans} already qualitatively establishes the spatiotemporal wavevector dynamics of the TML beam, with each spatiotemporal beam section only coupling to the fiber at a certain fiber angle due to the fiber coupling condition in Eq. (\ref{eq:coupling}).
To quantify the extent of the wavevector tilts present in the TML beam and to give a visual representation of the entire TML beam, first the input beams spatiotemporal intensity distribution is retrieved by superimposing the transmitted intensity distributions $I_{\alpha}(\Tilde{y},t)$ (see examples in Fig. \ref{fig:yscans} a)-c)) of all fiber angles $\alpha$:
\begin{equation}
    I_{\mathrm{total}}(\Tilde{y},t)=\sum_{\alpha} I_{\alpha}(\Tilde{y},t).
\end{equation}
Similarly, the beams wavevector angles are retrieved by calculating a weighted average of the spatiotemporal angular distributions, where the weights for each angle are given by the transmitted intensities $I_{\alpha}(\Tilde{y},t)$:
\begin{equation}
    \Bar{\alpha}(\Tilde{y},t)= \frac{\sum_{\alpha}I_{\alpha}(\Tilde{y},t) \alpha}{\sum_{\alpha}I_{\alpha}(\Tilde{y},t)}.
\end{equation}
In order to give a complete overview of the TML beam dynamics, the resulting intensity and wavevector distributions are shown in Fig. \ref{fig:fiber_scan_results1} a1) and b1), respectively.
\begin{figure}[h]
\centering
\includegraphics[width=1\linewidth]{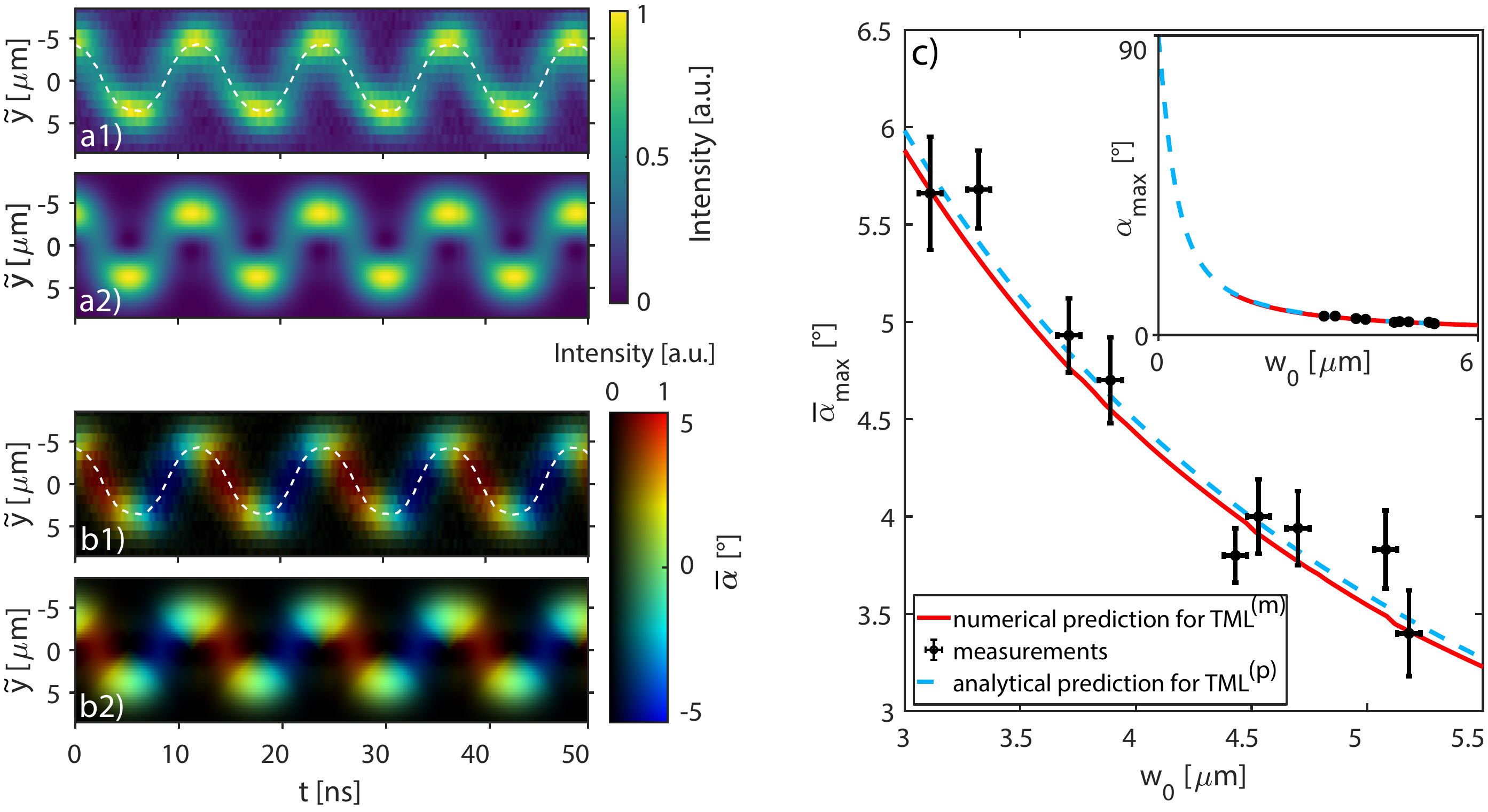}
\caption{a1) Retrieved TML beam with the same fundamental mode width from angular and spatial fiber scans as shown in Fig. \ref{fig:yscans}. a2) Calculated spatiotemporal intensity oscillation of TML beam with $\Bar{n}=0.86$ and a fundamental mode radius of $w_0=\qty{3.86}{\um}$. b1) Wavevector tilt angles in the retrieved beam. b2) Calculated wavevector tilt angles for a $\mathrm{TML}^{(\mathrm{m})}$ beam. Wavevector angles are colorcoded, while the respective intensities are coded in brightness. c) Black crosses: Maximum experimentally measured wavevector angle $\alpha_{\mathrm{max}}$ along the retrieved beam's center of mass as shown with a dashed white line in a1) and b1) as a function of fundamental mode radius $w_0$. The modeled dependence for the $\mathrm{TML}^{(\mathrm{p})}$ beam with average mode order $\Bar{n}=0.86$ according to Eq. (\ref{eq:max_alpha}) is shown as a dashed blue line. Shown with a solid red line, $\alpha_{\mathrm{max}}(w_0)$ for the $\mathrm{TML}^{(\mathrm{m})}$ beam was also numerically calculated. Inset: Wavevector angles over a larger range of $w_0$.}
\label{fig:fiber_scan_results1}
\end{figure}
The resulting intensity distribution in Fig. \ref{fig:fiber_scan_results1} a1) is in good agreement with a numerically calculated beam based on the modal weights and phases of the $\mathrm{TML}^{(\mathrm{m})}$ beam (Fig. \ref{fig:fiber_scan_results1} a2)).
The wavevector angles as shown in Fig. \ref{fig:fiber_scan_results1} b1) and b2) for measured and numerically calculated wavevector angles, respectively, show the oscillating spatiotemporal behavior of the TML beam's wavevector. Around the return points, the wavevector angle is close to zero, while the central beam sections exhibit the highest wavevector angle. Therefore, the wavevector direction oscillates with the frequency of the transverse mode separation $\Omega$, just as the spatiotemporal intensity oscillation does.\\
The spatiotemporal wavevector angles shown here were measured for a fundamental mode radius of $w_0=\qty{3.86}{\um}$. According to Eq. (\ref{eq:max_alpha}), the maximum wavevector angles in the TML beam is expected to change with changing beam sizes, which could be accomplished by moving lens L1 and thus changing the focal length of the lens combination (cf. Fig. \ref{fig:setup}).
To examine the impact of varying beam size on the wavevectors, the maximum wavevector angle $\alpha_{\mathrm{max}}$, found along the retrieved beam's center of mass (white dashed line in Fig. \ref{fig:fiber_scan_results1} b1)), was compared for different beam sizes. The thus retrieved maximum wavevector angles are shown as black dots in Fig. \ref{fig:fiber_scan_results1} c), with the error bars given by the standard deviation of measured fundamental mode size $w_0$ and maximum wavevector angles $\alpha_{\mathrm{max}}$, respectively.
The maximum wavevector angle increased with decreasing beam size in close agreement with Eq. (\ref{eq:max_alpha}) shown by a dashed blue line in Fig. \ref{fig:fiber_scan_results1} c), which represents the maximum wavevector angle in a $\mathrm{TML}^{(\mathrm{p})}$ beam with $\Bar{n}=0.86$. Represented with a solid red line in Fig. \ref{fig:fiber_scan_results1} c), the maximum wavevector angle of the TML beam as reconstructed by the SPGD algorithm (Fig. \ref{fig:beam_charac} c)) was numerically retrieved.
By changing beam size, wavevector angles from $3.4^{\circ}$ for $w_0=\qty{5.2}{\um}$ up to $5.7^{\circ}$ for $w_0=\qty{3.1}{\um}$ were accomplished. Larger wavevector angles in TML beams would occur when employing tighter focussing of the beam as can be seen in the inset of Fig. \ref{fig:fiber_scan_results1} c).\\
Nevertheless, even the experimentally measured wavevector angles are sufficiently large to impact phase-sensitive processes like optical fiber coupling as can be seen in Fig. \ref{fig:yscans} c2), where the TML beam's center could not be coupled into the SMF.
But also nonlinear frequency conversion processes, that rely on phase matching to achieve high conversion efficiencies, can be inhibited by  wavevector misalignment. 
In general, nonlinear processes usually demand tight focussing to increase local intensities, thus expediating the problem for TML beams as shown in the inset of Fig. \ref{fig:fiber_scan_results1} c). Thus, we conclude that wavefronts of TML beams need to be considered carefully when applying them to phase-sensitive processes, especially when small beam sizes are required.\\
\section{Summary}
The intensity profile of transverse mode-locked (TML) beams has been known to periodically scan space perpendicular to their propagation axis. On the other hand, the phase dynamics have not been characterized to the same extent. By investigating the phase terms of TML beams, their wavevector was shown to be oscillating spatiotemporally.\\
This wavevector oscillation was also verified experimentally by exploiting the fact that coupling of a light field to a single-mode fiber can only occur when both intensity as well as phase are matched to the fiber's eigenmode. The measured transmissions for the different fiber angles and positions were used to reconstruct the average wavevector angles, showing the theoretically predicted spatiotemporal wavevector tilt. Additionally, it was shown that with decreasing beam width, the wavevectors in such beams become increasingly tilted.\\
By exploiting the fiber coupling condition for the characterization of the TML beam, the impact of wavevector tilt on fiber coupling is immediately becoming obvious, with fiber transmissions being heavily diminished by the wavevector mismatch. However, the significance of wavevector direction holds true also for other phase-sensitive processes, like frequency conversion in nonlinear media, where beams are usually strongly focused to increase local intensity. In any case, the here-revealed spatiotemporal phase features need to be taken into account in any phase-sensitive application involving TML beams.\\
The findings about the spatiotemporally oscillating wavevectors also hold true for different types of TML beams, such as Laguerre-Gaussian-based TML beams \cite{Schepers2021} or the proposed two dimensional plane-scanning TML beams based on Lissajous curves \cite{Zwilich2023}, although the wavevector oscillation trajectories would be different in each case.
\begin{backmatter}
\bmsection{Funding} Open Access Publication Fund of the University of Münster.
\bmsection{Acknowledgments} We acknowledge support from the Open Access Publication Fund of the University of Münster.
\bmsection{Disclosures} The authors declare no conflicts of interest.
\bmsection{Data availability} Data underlying the results presented in this paper are not publicly available at this time but may be obtained from the authors upon reasonable request.
\end{backmatter}
\bibliography{bib2}
\end{document}